\documentclass[graybox,natbib,nosecnum]{svmult}
\bibpunct{(}{)}{;}{a}{}{,} 

\pdfoutput=1   

\usepackage{mathptmx}       
\usepackage{helvet}         
\usepackage{courier}        
\usepackage{type1cm}        

\usepackage{makeidx}         
\usepackage{graphicx}        
\usepackage{multicol}        
\usepackage[bottom]{footmisc}
\usepackage[normalem]{ulem}	
\usepackage{hyperref}  

\usepackage{soul}   

\newcommand{\ms}{m\,s$^{-1}$}
\newcommand{\cms}{cm\,s$^{-1}$}
\newcommand{\arcsec}{''}
\newcommand{\mearth}{$M_\oplus$}
\newcommand{\msun}{$M_\odot$}

\makeindex             

\begin{document}

\title*{ESPRESSO on VLT: An Instrument for Exoplanet Research}
\author{Jonay I. Gonz\'alez Hern\'andez, Francesco Pepe, Paolo Molaro and Nuno Santos}
\institute{
Jonay I. Gonz\'alez Hern\'andez \at Instituto de Astrof{\'\i} sica de Canarias, 
V{\'i}a L\'actea S/N, E-38205 La Laguna, Tenerife, Spain\\ \email{jonay.gonzalez@iac.es} \\
Universidad de La Laguna (ULL), Dpto. Astrof{\'\i}sica, E-38206 La Laguna, Tenerife, Spain  
\and 
Francesco Pepe \at D\'epartement d'astronomie de l'Universit\'e de Gen\`eve, 
Chemin Pegasi 51, 1290 Versoix, Switzerland \\ 
\email{Francesco.Pepe@unige.ch}
\and 
Paolo Molaro \at INAF-Osservatorio Astronomico di Trieste, Via G.B. Tiepolo 11, 34143 
Trieste, Italy\\
Institute of Fundamental Physics of the Universe, Via Beirut 2, 34151 Trieste, Italy\\  
\email{paolo.molaro@inaf.it}
\and 
Nuno Santos \at Instituto de Astrof{\'\i}sica e Ci\^encias do Espa\c{c}o, Universidade do 
Porto, CAUP, Rua das Estrelas, 4150-762 Porto, Portugal\\
Departamento de F{\'\i}sica e Astronomia, Faculdade de Ci\^encias, Universidade do 
Porto, Rua do Campo Alegre, 4169-007 Porto, Portugal\\
\email{nuno.santos@astro.up.pt}}
%
%
\maketitle

\abstract{ESPRESSO ({E}chelle {SP}ectrograph for {R}ocky {E}xoplanets and 
{S}table {S}pectroscopic {O}bservations) is a VLT ultra-stable high 
resolution spectrograph installed at ESO's Paranal Observatory in Chile at the end of 2017
and that started regular operations in October 2018. 
The spectrograph is located at the VLT Combined-Coud\'e Laboratory and is
able to operate with one or (simultaneously) four 8.2\,m Unit 
Telescopes (UTs) through four optical Coud\'e trains. 
Combining efficiency and extreme spectroscopic precision, ESPRESSO has 
demonstrated to gain about two magnitudes with respect to its predecessor HARPS. 
ESPRESSO has improved the instrumental radial-velocity precision getting close 
to the aimed 10 \cms~level, thus opening the possibility to explore new frontiers 
in the search for Earth-mass exoplanets in the habitable zone of quiet, nearby G to 
M-dwarfs. ESPRESSO will be certainly an important development step towards  
high-precision ultra-stable spectrographs on the next generation of giant telescopes 
such as the ELT.}

\section{Introduction}

During the last decades the exoplanet science has become one of the most exciting 
research fields in modern astrophysics. The doppler spectroscopy or radial velocity 
(RV) technique (based on the determination of the projected velocity of stars in the 
direction of the line of sight) was the earliest method delivering the first detection 
of low-mass companions~\citep[e.g. HD114762 b --][]{lat89}. 
This technique provided in 1995 the first discovery of a Jupiter-mass exoplanet 
orbiting the solar-type star 51~Pegasi~\citep{may95Natur}. 
After the discovery of 51 Peg b, direct imaging, microlensing and specially transit 
searches both from the ground and space~\citep{lis14Natur}, together with the RV 
technique~\citep{may14Natur}, have produced an increasing number of detections of 
planets and planetary systems. 
The number of confirmed exoplanets discovered as of December 2023 is 
about 5557, among which $\sim 4146$ exoplanets have been detected using the 
transit technique, mostly from the 
CoRoT\footnote{Convection, Rotation and planetary Transits (CoRoT)}~\citep{bar06}, 
Kepler~\citep{bor09}, 
K2~\citep{how14}  and TESS\footnote{Transiting Exoplanet Survey Satellite 
(TESS)}~\citep{ric15} space missions, with 37, 2778, 548, and 
410 confirmed planets, respectively. 
About 1071 exoplanets have been discovered using the doppler technique on 
high-resolution spectrographs. Combined observations from both the transit and the 
RV techniques have yielded a significant   
statistical contribution to our understanding of exoplanet population, such as the 
frequency of planets~\citep[e.g.][]{may11,how12,kun20,rib23}, for different host
spectral types and at different orbital distances including the habitable 
zone (i.e. in orbits where water is retained in liquid form on the planet surface).

Now we know there is wide variety of planetary systems with different planetary 
masses, sizes, orbital periods, eccentricities and different configurations of 
mass/size-distances to the host stars.
Most of the known planets appear to be in planetary systems and a significant fraction 
in multiple planet systems. Examples of complex planetary systems as the Solar 
System are the 7-planet system orbiting the 
G dwarf HD10180, detected using the RV technique with six confirmed 
planets in the mass range $11-65$~\mearth\ and one planet candidate with 
a mass as small as $1.5$~\mearth\ \citep{lov11}, or the more recent discovery
using transit technique from the ground of also seven planets with similar size as that 
of the Earth orbiting the late M dwarf TRAPPIST-1~\citep{gil17Natur}. 
More recently, the discovery and characterization of a 6-planet  
system in a chain of Laplace resonances orbiting the late K-dwarf 
TOI-178~\citep{lel21} has shown the great synnergy between  
the TESS and CHEOPS\footnote{CHaracterising ExOPlanets Satellite 
(CHEOPS)}~\citep{ben21} space missions, and the 
ground-based ESPRESSO@VLT facility.
The discoveries have required intensive and long-term monitoring of stars and 
the continuous development of astronomical instrumentation~\citep{pep14Natur}. 
This together with significant improvement in the observational strategies has 
made exoplanet science quickly evolve from the discovery of giant planets even 
more massive than Jupiter to rocky planets with similar mass as that of the Earth. 
This is demonstrated by the discoveries of Kepler 78~b~\citep{pep13Natur}, 
a planet with similar density as that of the Earth, and Proxima 
Centauri~b~\citep{ang16Natur}, a planet with similar minimum mass as that 
of the Earth in the habitable zone of the closest star to the Sun. 
ESPRESSO has allowed us to go beyond the Earth-mass frontier with the discovery of 
some new sub-Earth mass planets such as Proxima Centauri~d~\citep{sua20,far22}, 
the lowest mass exoplanet ever detected with the doppler technique, with a mass 
of $0.26\pm 0.05$~\mearth,  just about twice the mass of Mars, at an orbital period 
of $5.12\pm0.04$~days (0.029 AU from the star), with a RV semiamplitude 
of $39\pm 7$~\cms.

Coupling the transit with the RV technique allows us to derive the mass and radius, 
and therefore to compute bulk density of exoplanets, a first step towards 
characterization of exoplanets. This in combination with detailed theoretical 
modelling allows us to get some insights about the bulk composition of some 
exoplanets and how they compared 
with the Earth composition~\citep[see][and references therein]{may14Natur,lis14Natur}. 
ESPRESSO is breaking the frontier of characterizing planets smaller than the 
Earth, as e.g. the two super-Mercuries, less-massive but denser than the Earth, 
together with three super-Earths detected in the 5-planet system HD 23472~\citep{bar22}.

The next step is to charaterize the atmospheres of exoplanets, using e.g 
high-resolution spectroscopy via transmission with a technique applied to 
some transiting giant planets to detect Na and CO in their atmospheres~\citep{cha02,sne08,sne10,wyt15},
or via spectroscopic detection of reflected light of exoplanets~\citep{cha99,mar15}.
In spite of the difficulty of this type of analysis, ESPRESSO has managed to achieve remarkable results, demonstrating for instance the presence of rare species in the atmospheres of certain exoplanets, 
such as barium and lithium in the atmosphere of WASP-121 b~\citep{bor21,aze22,sei23}, and 
WASP-76 b~\citep{ehr20,bor21,tab21,sei21,aze22}.

Most of the known transiting exoplanets (namely those discovered by the Kepler 
space telescope) orbit faint targets, which make it difficult to study their 
atmospheres with the current facilities. 
The recently found transiting super-Earth like planet in the habitable 
zone of the faint M dwarf LHS~1140~\citep{dit17Natur}, discovered with 
MEarth~\citep{nut08} and HARPS~\citep{may03}, and later 
confirmed with TESS and ESPRESSO~\citep{lil20,cad23},
is an example of the need for larger telescopes equipped with 
high-resolution, high-precision spectrographs to investigate the atmospheres of such 
interesting exoplanets. 

In particular, the difficulty raises when trying to find transiting Earth-like planets 
orbiting at the habitable zones of bright host stars.
It appears quite necessary to design new techniques able to study the 
atmospheres of exoplanets by using the reflected light~\citep{sne14}, 
combining a high-contrast adaptive optics (AO) system with a 
high-resolution spectrograph to overcome the tiny planet-to-star flux ratio in two 
stages. 
First, the AO system should spatially resolve the planet from the host star enhancing
the planet-to-star contrast at the planet location and second, the light beam at the 
planet location passes through the high-resolution spectrographs. This technique has 
been applied to model the possible planetary atmospheric signal of Proxima Cen b 
by coupling the SPHERE high-contrast imager~\citep{beu08} and the ESPRESSO 
spectrograph at VLT~\citep{lov17}. 
A new instrument, RISTRETTO, for the VLT, that 
will provide a visible high-resolution spectrograph fed by an extreme adaptive 
optics (XAO) is being designed with the goal of detection and atmospheric 
characterization of exoplanets in reflected light, in particular, the temperate rocky 
planet Proxima b~\citep{lov22}.

The new generation of 20-40m giant telescopes such as ELT equipped with 
sophisticated AO systems and with stable high-precision high-resolution 
spectrographs will possibly be able to study the atmospheres of rocky exoplanets in 
the habitable zone~\citep{sne15}. 
The new optical and near-infrared ultrastable spectrograph 
ANDES\footnote{ArmazoNes high Dispersion Echelle Spectrograph (ANDES)} is in 
design phase~\citep{mar22} for the ELT, and will be able to characterize the 
atmospheres of Earth like planets orbiting nearby stars in the habitable zone both 
in transmission and reflected-light spectroscopy to search for 
biosignatures~\citep{pal23}.

The need of instrumentation for exoplanet science on the 8.2-m Very Large Telescopes 
(VLT) was highlighted in the ESO (European Southern Observatory)-ESA(European 
Space Agency) working report on extrasolar planets. 
In October 2007 the ESO Science Advisory Committee recommended the 
development of new second-generation VLT instrumentation, and later 
endorsed by the ESO Council. Among those instruments, ESPRESSO, a high-resolution 
ultra-stable spectrograph for the VLT combined-Coud\'e focus, was proposed.
The ESPRESSO ({E}chelle {SP}ectrograph for {R}ocky {E}xoplanets and {S}table 
{S}pectroscopic {O}bservations) project started with the kick-off meeting in 
February 2011. The ESPRESSO consortium is composed of: Observatoire 
Astronomique de l'Universit\'e de Gen\'eve (project head, Switzerland); 
Instituto de Astrof{\'\i}sica e Ci\^encias do Espa\c{c}o/Universidade 
de Porto and Universidade de Lisboa (Portugal); INAF-Osservatorio Astronomico 
di Brera (Italy); INAF-Osservatorio Astronomico di Trieste (Italy); Instituto de 
Astrof{\'\i}sica de Canarias (Spain); Physikalisches Institut der Universit\"at 
Bern (Switzerland). 
ESO participates to the ESPRESSO project as Associated Partner.
The ESPRESSO instrument was commissioned at the VLT in 
Paranal Observatory in the fall 2017 and offered to the community by 2018, 
starting regular operations in October 2018.

\section{Exoplanet science with ESPRESSO}

The main scientific drivers of ESPRESSO are:
(i) search for rocky planets; and
(ii) measure the variation of physical constant. 
We refer to \citet{pep14AN, pep21} for details on (ii), as well as very 
challenging additional scientific topics that ESPRESSO will address, as e.g. the use 
of the ESPRESSO RV stability to investigate long-period binarity in extremely iron-poor 
stars~\citep{agu22,agu23,mol23}.

The ESPRESSO Guaranteed Time Observations (GTO)~\citep{pep21} were executed between 
October 2018 and September 2023, and were organized in four main groups: 
(i) WG1: blind search for exoplanets; 
(ii) WG2: atmospheric characterization of giant exoplanets via transmission spectroscopy; 
(iii) WG3: mass measurements of transiting exoplanets discovered by K2 and TESS; 
(iv) WG4: fundamental constants from QSO observations. 

\begin{figure}
\vskip-2mm
\includegraphics[width=10.5cm]{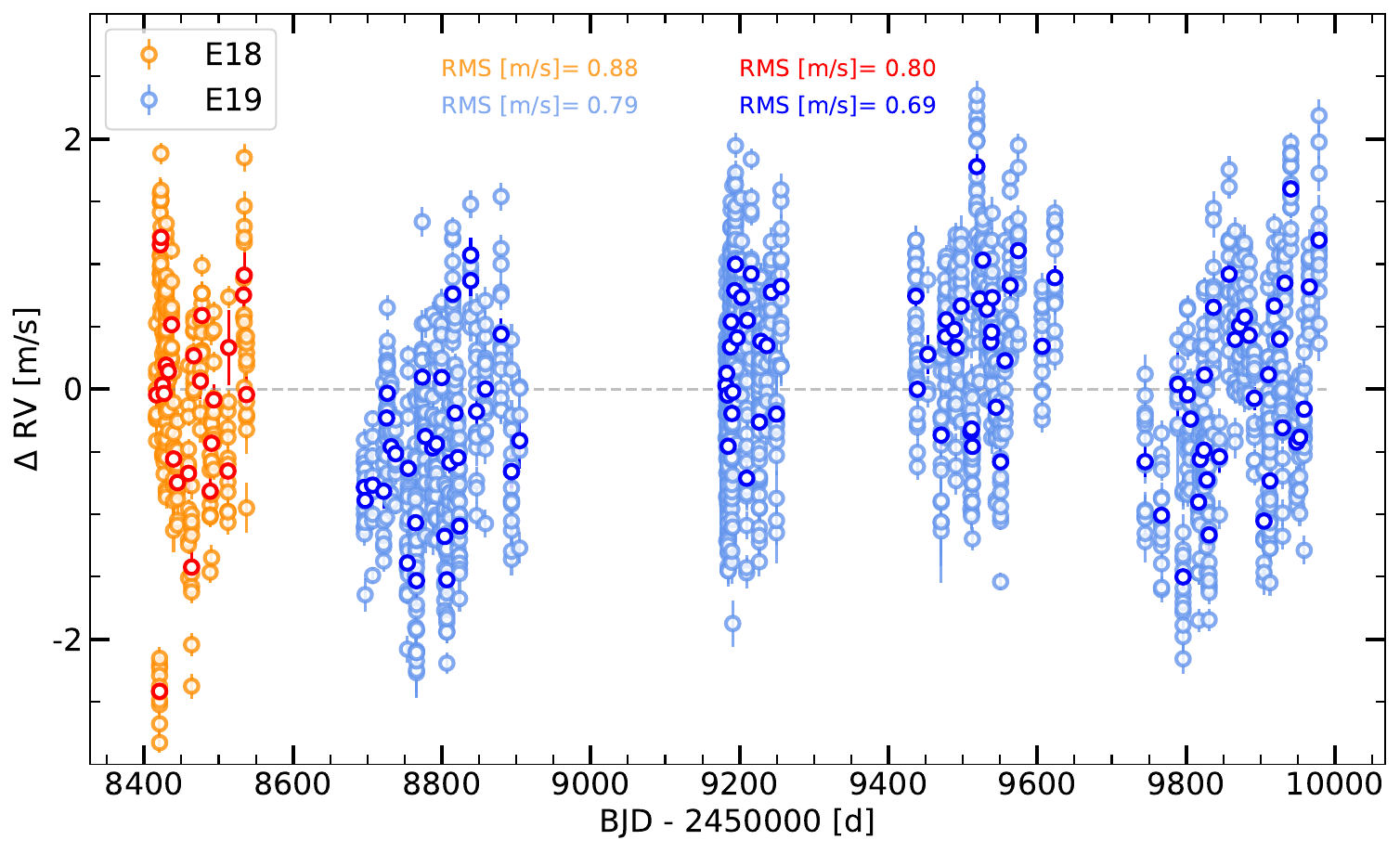}
\caption{
$4.4$-years ESPRESSO RV data of the late G-type dwarf star $\tau$ Ceti
before (light orange) and after (light blue) the intervention in June 2019. 
The overall dispersion is below 90~ \cms. The nightly-binning RV data before (red) 
and after (blue) show rms values  below 80 and 70~\cms, respectively.  
The significant long-term trend in the data is remarkable, thus demonstrating the 
outstanding precision of the ESPRESSO spectrograph. 
}
\label{figtaucet}
\end{figure}

One of the main scientific topics in the next decades is the search and charaterization
of terrestrial planets in the habitable zone of their host stars, and one of the main 
drivers of the new generation of extremely large telescopes is the detection of their 
atmospheres~\citep[see e.g.][]{mar16, pal23}.
At the end of the 90's and the following decade, the monitoring of stars using 
high-resolution spectroscopy yielded many detections of Jupiter like planets. 
Only after 2003,  the HARPS spectrograph installed at the 3.6-m ESO telescope in 
La Silla Observatory (Chile) opened a new window on the domain of 
Neptune and super-Earth like planets (see the {\it Review} by Pepe et al. in 
this {\it HandBook of Exoplanets},~\citet{pep18haex}).
This instrument (with a resolving power of $R\sim115,000$) is contained inside a 
vacuum vessel and uses a simultaneous calibration reference that allows to correct 
for (small) instrumental drifts due to (also small) changes of temperature 
and pressure, providing an extreme long-term RV precision below 
the 1~\ms~\citep[see Fig. 1 in][]{pep14AN}. 
The ESPRESSO instrument has confirmed the late G-type dwarf star $\tau$ Ceti 
as a RV standard where the  roughly 1~\ms RV variation is mostly caused by the 
stellar activity~(see Fig.~\ref{figtaucet}). 
Already in 2004 the discovery of a super-Earth with a minimum mass of about only
10~\mearth\ orbiting the G-type star $\mu$~Arae~\citep{san04} demonstrated 
the impressive capabilities of the HARPS instrument. Since then, and thanks
to its precision, HARPS has provided the discovery of most of the sub-Neptunian
mass planets with masses down to a few Earth-masses inducing RV semi-amplitudes
as low as 0.5 \ms~\citep[see e.g.][]{pep11}. 

\begin{figure}
\vskip-2mm
\includegraphics[width=11.5cm,angle=0]{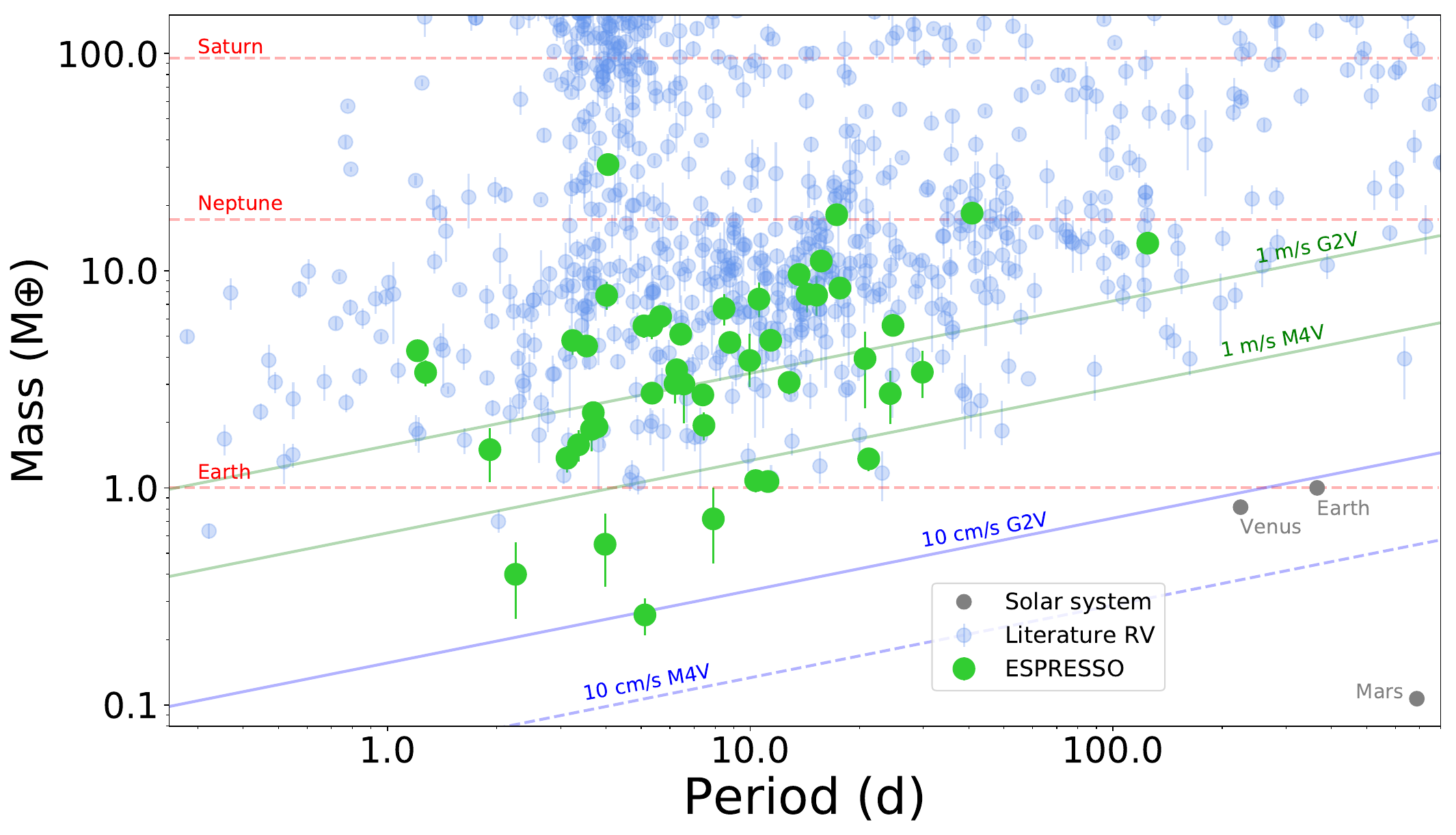}
\includegraphics[width=11.5cm,angle=0]{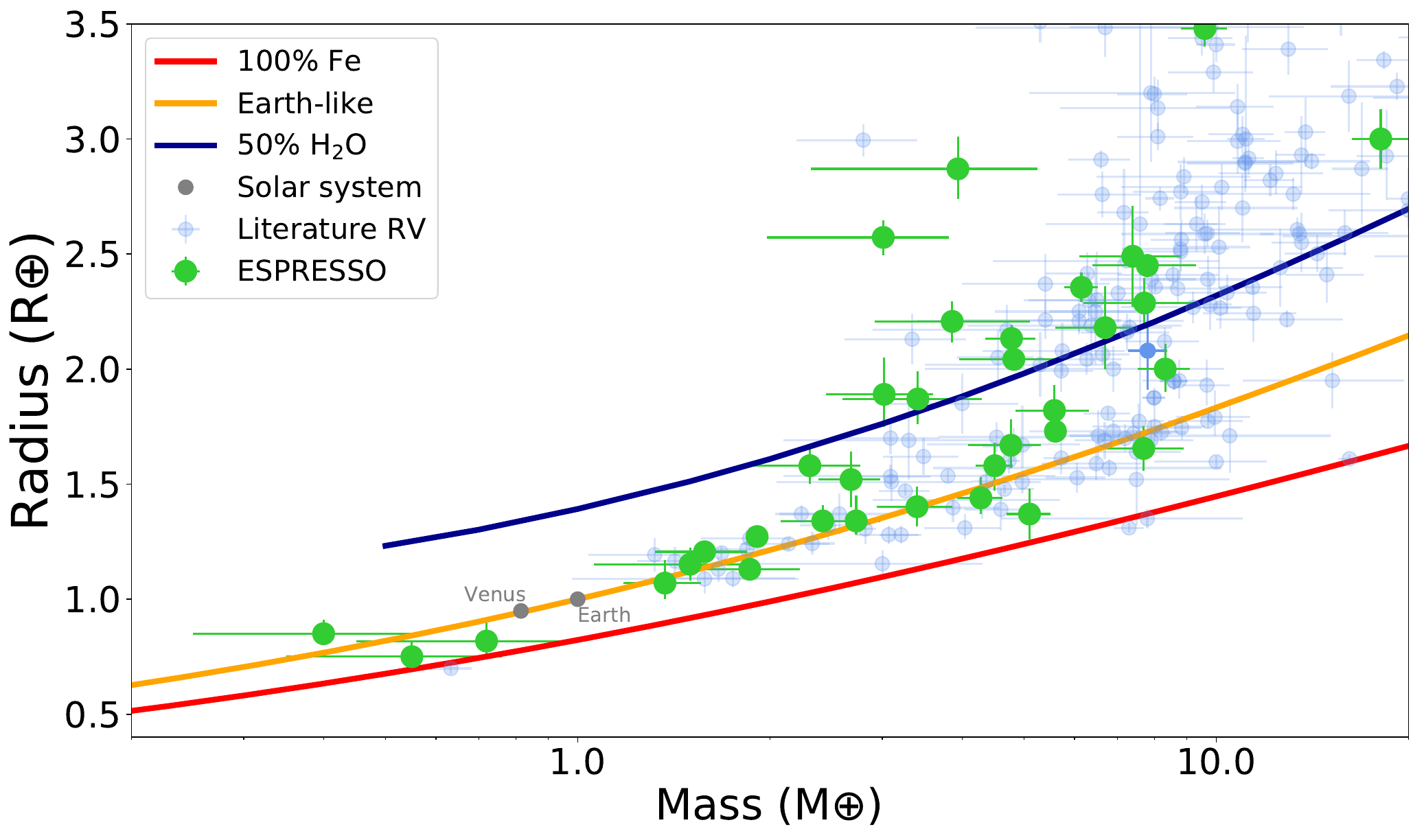}
\caption{ 
{\it Top:} Minimum mass vs orbital period diagram for known planets from the 
NASA Exoplanet archive as of December 2023 orbiting solar-type stars, together 
with those discovered and confirmed using ESPRESSO (green circles).
Inclined solid and dashed lines show the RV semiamplitude of planets orbiting 
a late M dwarf star with 0.25~\msun\ (green line) and a G dwarf star with 
1~\msun\ star (blue line) assuming a RV semiamplitude of 1 \ms 
and 10 \cms, respectively, and null eccentricity.
Planets of the solar system (grey circles) are also labeled.  
{\it Bottom:} Radius vs mass diagram for known transiting planets. ESPRESSO 
results are highlighted using green circles. Mass-radius models are also displayed 
for three different bulk's planet compositions: rocky planets with 100\% Fe (red) 
and Earth-like rocky planets with 33\% Fe (orange) and water worlds with 
50\% H$_2$O (blue)~\citep{zen19}. 
}
\label{figplanet}
\end{figure}

The ESPRESSO instrument has improved a step further, aiming at achieving a RV precision 
of 10~\cms which is crucial in the path of detecting terrestrial planets in the
habitable zone of host stars of different spectral types, in particular for G, and 
K dwarfs. 
The Earth induces radial velocity variations in the Sun of $\sim 9$~\cms, in comparison 
with RV signal of 12~\ms induced by Jupiter. However, an Earth-mass exoplanet
orbiting a M5V star in the habitable zone would cause a gravitational pull equivalent
to $\sim1.3$~\ms which, with the current instrument capabilities limited to 
$\sim 1$~\ms, can be detected~\citep{bon13M,ang16Natur}.
Thus, the quest for low-mass planets in the habitable zone has favoured RV studies 
on M dwarf in the last decade with numerous low-mass planets 
detections~\citep[see e.g.][]{bon13M,bon13pl,ang13,aff16,ast17,sua17pl,sua17plhz}.
Today, several tens of planets with minimum masses below 10~\mearth\ have 
been discovered.
Most of them have been detected orbiting cool dwarfs less massive than the 
Sun (see Fig.~\ref{figplanet}), using HARPS and HARPS-N~\citep[at 3.6-m TNG telescope 
located in the {\it Observatorio del Roque de los Muchachos}, La Palma, Spain;][]{cos12} 
spectrographs. 
New dedicated instruments aiming for planet search around M dwarfs operating in the 
near infrared are already running such as CARMENES~\citep[at the 3.5-m telescope in the 
{\it Observatorio de Calar Alto}, Spain;][]{qui16}, NIRPS~\citep[at the 3.6-m ESO 
telescope in La Silla (Chile), in operations together with HARPS since 
April 2023;][]{bou17}, and SPIRou~\citep[at the CHFT telescope in Mauna 
Kea (Hawaii, USA);][]{don20}.

A larger telescope size provides a lower photon noise level on Doppler signal for the 
same exposure time. ESPRESSO at the 8.2-m VLT at the {\it Observatorio de Paranal} 
is expected to achieve the 10 \cms Doppler precision and long-term 
stability~\citep{pep21}. 
This will open the possibility to search for Earth-mass planets at different orbital 
distances, including the habitable zones of solar-type stars. A carefully selected 
sample of non-active, non-rotating, quiet G to M dwarf will allow to explore this 
new domain.
In the top panel of Fig.~\ref{figplanet} we display the known planets from the 
NASA Exoplanet archive as of December 2023 at different orbital periods around stars.
So far, most of the low-mass planets have been discovered around M dwarfs where the 
RV planetary signals are stronger. We have highlighted the ESPRESSO discoveries and 
confirmations.
The larger telescope mirror of VLT and the RV precision provided by ESPRESSO 
is allowing to access a larger sample of fainter stars of different spectral types. 
The recent discovery of two temperate Earth-mass planets in the habitable 
zone of the relatively faint M5.5V star ($m_V \sim 13.8$~mag, but brighter in 
the near infrared, 
$m_J\sim 8.3$~mag) GJ 1002 demonstrates the capabilities of the 
ESPRESSO instrument~\citep{sua23}.
During the execution of the ESPRESSO GTO, it has been possible to break 1~\ms 
limit and to go beyong the Earth-mass barrier, as e.g. with the discovery of 
the transiting multi-planet system L 98-59 with planet b of half of the mass of 
Venus~\citep{dem21}, 
as well as the aforementioned blind RV detection of Proxima d with almost twice 
the mass of Mars~\citep{far22}. 
In the bottom panel of Fig.~\ref{figplanet} we also display the mass-radius 
diagram of known transiting exoplanets, highlighting those confirmed using 
ESPRESSO, in comparison with models with different bulk's compositions.

However, stellar noise or jitter, which causes different radial velocity variations at 
different timescales and of different magnitude
\citep[see e.g.][]{saa98,san00,que01,boi11,dum11,rob14,raj15,sua17rv,dum18} still 
remains the main source of error and probably the strongest limitation towards 
the sub-\ms precision in the long term. 
Therefore, continuous investigation and modelling on these stellar effects at 
different timescales are required on the way towards finding rocky planets in 
the habitable zones of solar-type stars. One significant effort in this respect is 
the development of the PoET telescope~\footnote{http://poet.iastro.pt}, a solar 
telescope that will connect to ESPRESSO and allow to obtain ultra-high resolution 
spectra of resolved solar regions. The data will allow to understand in unique 
detail the different sources of stellar noise in both Doppler radial velocity 
measurements and transmission spectroscopy of planets orbiting sun-like stars.

\begin{figure}
\center{
\includegraphics[width=11.5cm,angle=0]{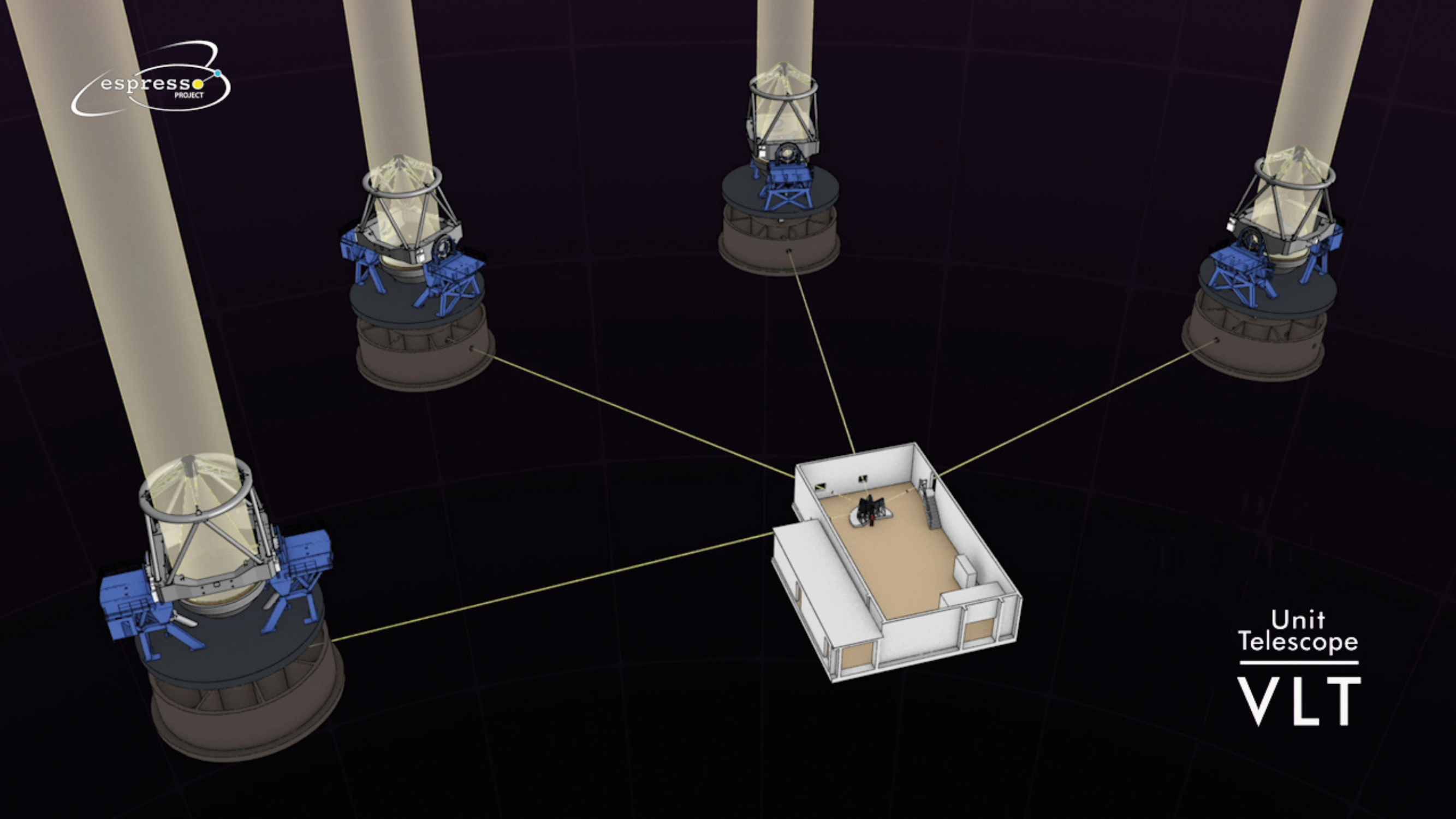}}
\caption{Schematic view of the four 8.2m Unit Telescopes of the VLT (multi-UT mode)
feeding, through the Coud\'e train, the Front-End unit of the ESPRESSO spectrograph 
located in the Combined Coud\'e Laboratory. 
}
\label{fig4ut}
\end{figure}

The discovery of a new and large population of Earth-mass exoplanets orbiting solar-type
stars will expand our knowledge of planet formation, and will also deliver new candidates
for follow-up observations using other techniques such as transit, astrometry, and 
Rossiter-McLaughlin (RM) effect. 
The detection with HARPS of the RM effect in occasion of the 2012 transit of Venus 
over the disk of the SUN provided a sort of preview of the kind of the physical 
information which we can hope to obtain by observing transits of exoplanets with a 
large telescope~\citep{mol13RM}.
ESPRESSO can also perform follow-up observations of ongoing and forthcoming 
transit surveys such as NGTS, MEarth from the ground or Kepler-K2, TESS and, in 
future PLATO\footnote{PLAnetary Transits and Oscillations of stars (PLATO)}~\citep{rau14,rau18}, from space. 

ESPRESSO will be possibly one of the rare instruments able to confirm 
long-period Earth-size transiting planets discovered by space missions like PLATO, 
which hopefully will provide Earth-size candidates transiting bright targets in their 
habitable zones. 

\begin{figure}
\includegraphics[width=5.5cm,angle=0]{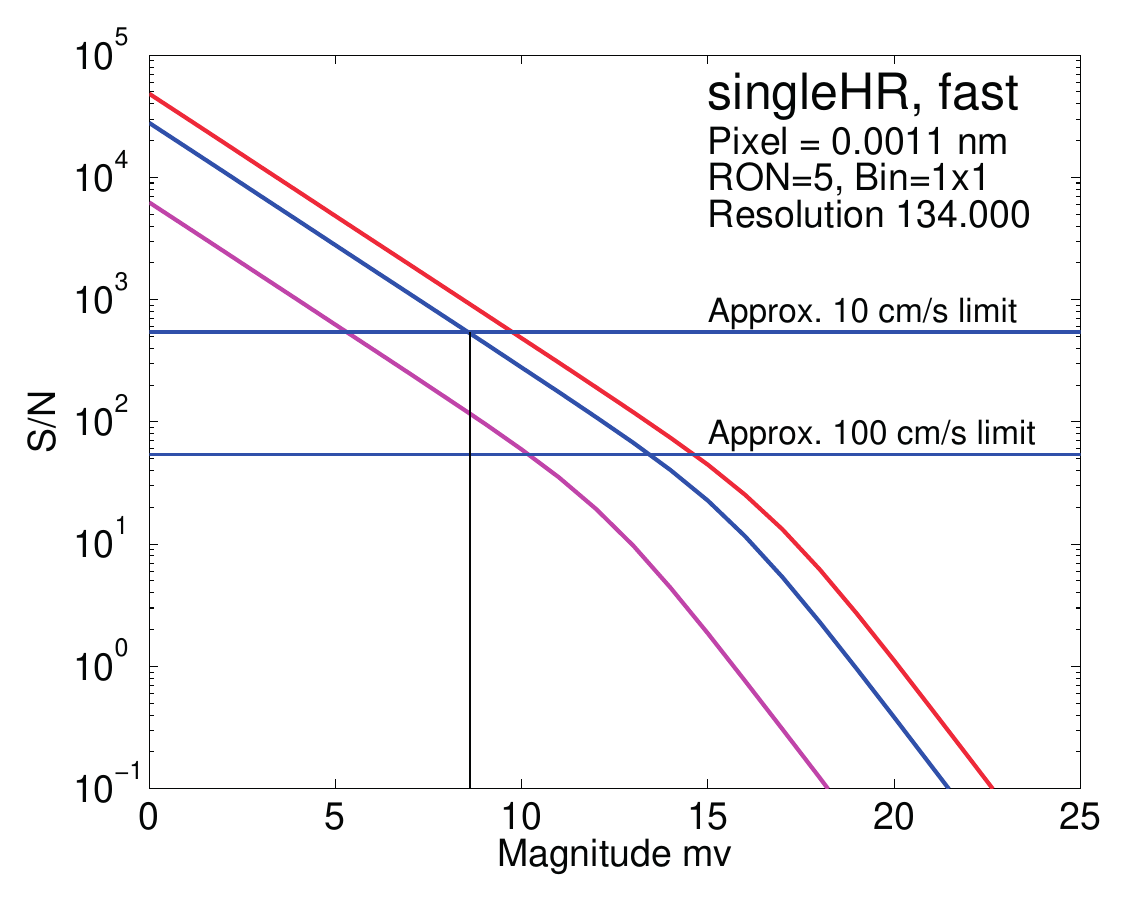}
\includegraphics[width=5.5cm,angle=0]{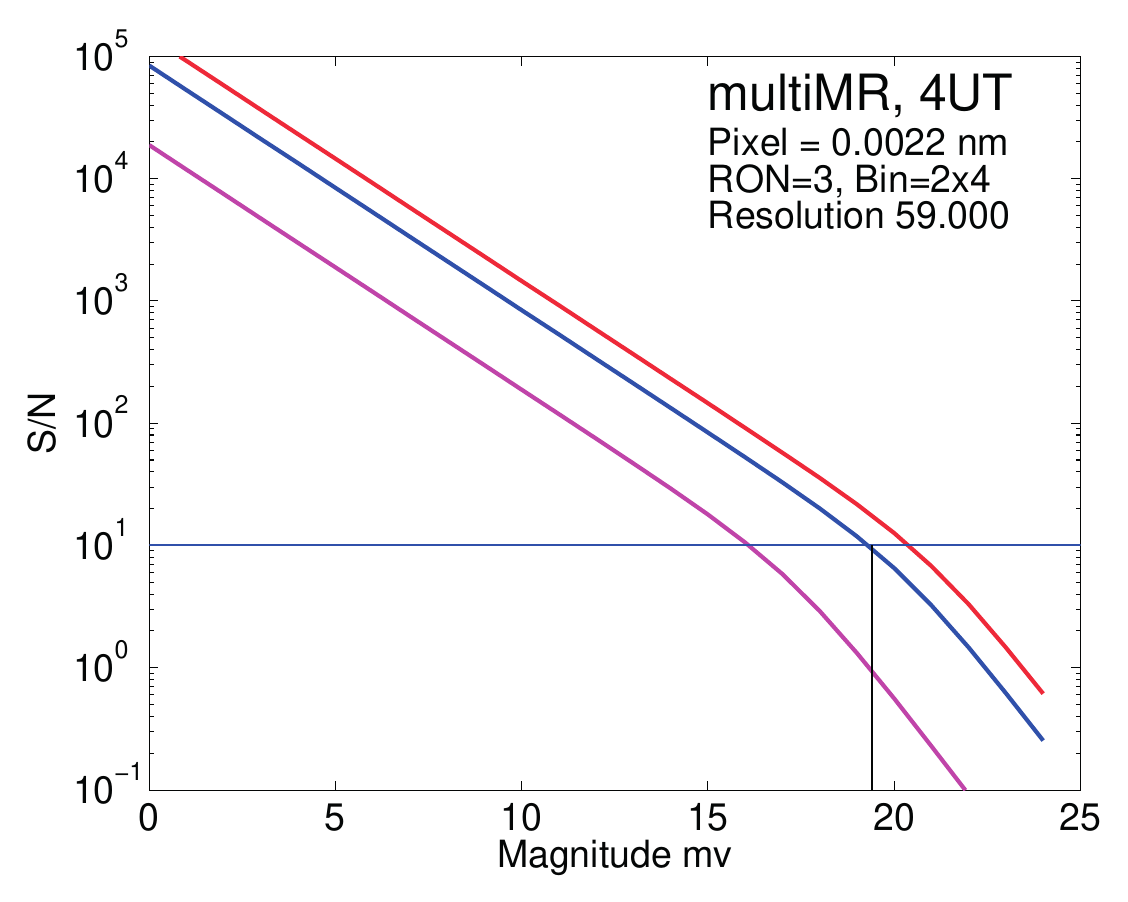}
\caption{Expected signal-to-noise ratio versus the stellar aparent visible magnitude
for the singleHR mode (left panel) and the multiMR mode (right panel). Red, blue, and
magenta curves indicate exposure times of 3600 s, 1200 s and 60 s, respectively.
This figure has been taken from \citet{pep14AN}, courtesy of F. Pepe.
}
\label{figstn}
\end{figure}

The high resolution and stability of ESPRESSO allows for atmospheric characterization 
of exoplanets of different mass and size using the transmission and 
possibly reflection spectroscopy~\citep{cha02,sne14,mar15,lov17,san20,cas21,cas22}. 
ESPRESSO GTO data have allowed to characterize in detail the atmosphere of 
the ultra-hot jupiter WASP-76 b~\citep{ehr20,sei21}, which has become a 
benchmark target in exoplanet atmospheres, getting deeper into the structure 
of the iron storm. In addition, ESPRESSO observations done using
1-UT and 4-UT modes have shown the extreme capabilities of 
ESPRESSO~\citep{bor21,sei23}, combining high-resolution, stability 
and the collecting area of the four 8.2m unit telescopes of the VLT to discover 
the presence of the species Ba+ in WASP-121 b and Li in both WASP-121 b 
and WASP-76 b \citep{bor21,aze22,tab21}.

\section{A ultra-stable high-resolution spectrograph for the VLT}

ESPRESSO is a fiber-fed, cross-dispersed, high-resolution \'echelle spectrograph 
located in the Combined Coud\'e Laboratory (CCL) at the incoherent focus, where a 
front-end unit can combine the light from up to four Unit Telescopes (UT) of the VLT.
The so-called Coud\'e train optical system feeds the light of each UT to the 
spectrograph. 
The sky light and the target enter the instrument simultaneously through two 
separate fibers, which form together the {\it slit} of the spectrograph. 
ESPRESSO, unlike any other ESO instrument, is able to receive the 
light from any of the four 8.2-m UTs, and is able to operate simultaneously with 
the light of either one UT or several UTs (see Fig.~\ref{fig4ut}). 
We refer to \citet{pep21} for updated ESPRESSO instrument details and 
commissioning results, that we only briefly will describe below.

\begin{figure}
\center{
\includegraphics[width=11.5cm,angle=0]{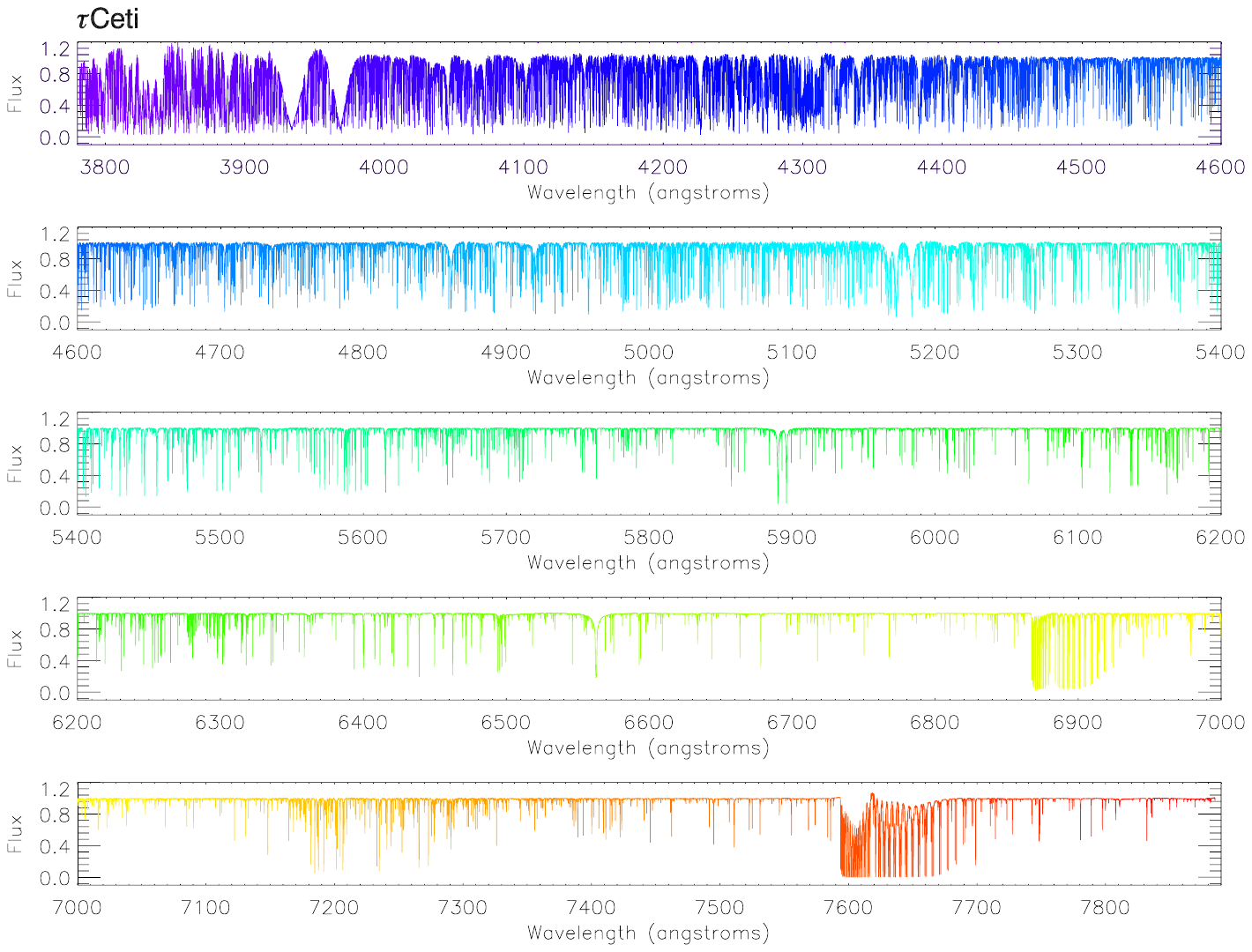}}
\caption{
Normalized, co-added ESPRESSO spectrum of $\tau$~Cet star obtained with 
the {\sc Star II} workflow of the data analysis software (DAS) of the ESPRESSO 
ESO DAS pipeline using 40 individual HR11 ESPRESSO spectra. 
This figure has been taken from \citet{pep21}.
}
\label{figspectcet}
\end{figure}

\subsection{Observing modes and performance}

The extreme precision of ESPRESSO is achieved based on well-known 
concepts provided by the HARPS experience. 
The light of one or several UTs is fed through the front-end unit into optical 
fibers that scramble the light and provide excellent illumination stability to the 
spectrograph. 
In order to improve light scrambling, non-circular but octogonal or square fiber 
shapes are used~\citep{cha12}.
The target fiber can be fed either with the light from the astronomical object 
or one of the calibration sources.
The reference fiber receives either sky light (faint source mode) or
calibration light (bright source mode). 
In the latter case, the simultaneous-reference technique successfully applied in 
HARPS and confirmed in ESPRESSO enables to track instrumental drifts down to
the \cms\ level.
In this mode the measurement is photon-noise limited and detector read-out noise
negligible. 
In the faint-source mode, instead, detector noise and sky background
may become significant. 
In this case, the second fiber allows to measure the sky background, whereas a 
slower read-out and high binning factor reduces the detector noise. 
ESPRESSO provides spectra in the wavelength range 378.2-788.7~nm with 
three instrumental modes: singleHR, singleUHR and 
multiMR, with each different detector binning  
(HR11, HR21, HR42, UHR11, MR42, MR84, see Table~\ref{tabmod}). 
They produce correspondingly different spatial and spectral sampling. 
The resolving power, instead, is essentially given and fixed by the 
instrument mode; it varies only slightly with binning. The HR and the MR modes 
are furthermore available with two different detector read-out modes 
(slow/high-gain and fast/low-gain) optimized for low and high-SNR 
measurements, respectively.

\begin{table}
\caption{Observing modes of ESPRESSO.}
\centering
\begin{tabular}{lcccccc}
\hline\noalign{\smallskip}
Obs. Mode     & HR11 (1UT)     & HR21 (1UT)     &  HR42 (1UT)    &  UHR11 (1UT)   & MR42(4UT)           & MR84(4UT)           \\
\hline\noalign{\smallskip}
Wave. range   & 380--790 nm    & 380--790 nm    & 380--790 nm    & 380--790 nm    & 380--790 nm         & 380--790 nm         \\
Resol. Power  & 138,000        & 138,000        & 130,000        & $>190$,000     & 72,500              & 70,000             \\
Aper. on Sky  & 1.0\arcsec     & 1.0\arcsec     & 1.0\arcsec     & 0.5\arcsec     & 4$\times$1.0\arcsec & 4$\times$1.0\arcsec \\
Spec. Samp.   & 4.5 pix        & 4.5 pix        & 2.25 pix       & 2.5 pix        & 5 pix               & 2.5 pix             \\
Spat. Samp.   & 2$\times$9 pix & 2$\times$4.5 pix & 2$\times$2.25 pix & 2$\times$5 pix & 2$\times$20 pix     & 2$\times$10 pix     \\
Sim. Ref.     & Yes (no sky)   & Yes (no sky)   & Yes (no sky)   & Yes (no sky)   & Yes (no sky)        & Yes (no sky)        \\
Sky Sub.      & Yes (no ref.)  & Yes (no ref.)  & Yes (no ref.)  & Yes (no ref.)  & Yes (no ref.)       & Yes (no ref.)       \\
Tot.  Eff.    & 10\,\%         & 10\,\%         & 10\,\%         & 5\,\%          & 10\,\%              & 10\,\%              \\
\hline
\end{tabular}
\label{tabmod}
\end{table}

The expected observational efficiency of ESPRESSO is shown in Fig.~\ref{figstn}, 
but the updated efficiency from ESPRESSO real observations is shown in 
Figs.~12-16 of \citet{pep21}.
In the singleHR mode (${R\sim 140\,000}$), we estimate SNR = 10 per 
extracted pixel in 20 minutes on a ${V = 16.3}$ star, or a SNR~$= 420$ on a 
${V = 8.6}$ star~\citep[see Fig.~16 in][]{pep21}.
We have estimated that at this resolution and a SNR~$\sim 420$ leads to 
$\sim 15-20$ \cms\ RV precision and reaching the 10 \cms\ level at SNR~$> 650$ 
for a non-rotating G8-K5 star~\citep[see Fig.~20 in][]{pep21}.
In the multiMR mode, at ${R\sim 70\,000}$, a  SNR~$\sim 10$  is 
achieved on a ${V = 19.4}$ star with an exposure of 20 minutes, 
a binning 4$\times$2 (MR42), and a slow read-out of the CCD.
In the following sections we briefly describe the several subsystems of 
the ESPRESSO project~\citep[see also][]{pep14AN,meg14,pep21}.

\subsection{The Coud\'e train}

\begin{figure}
\center{
\includegraphics[width=11.5cm,angle=0]{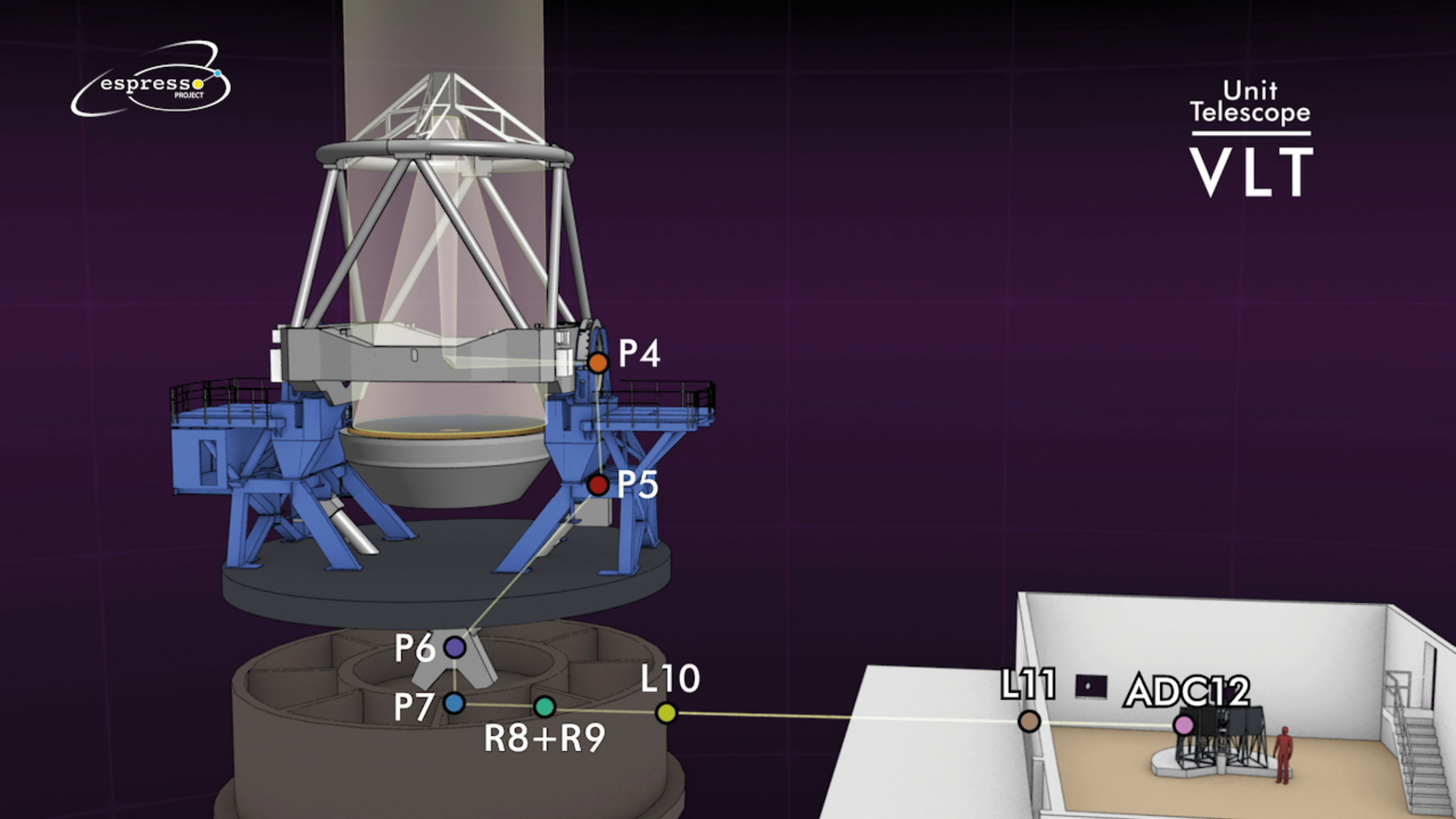}}
\caption{Schematic view of the Coud\'e train of ESPRESSO with the optical path 
through the telescope to the Combined Coud\'e Laboratory. 
The optical elements are highlighted at different positions along the path 
from the Nasmyth-B platform to the CCL with P (prism), R (mirror) and L (lens).
}
\label{figctrain}
\end{figure}

The four VLT telescope are connected to the CCL through four tunnels with a length 
that goes from 48 to 69 meters (see Fig.~\ref{fig4ut}). 
A trade-off analysis among the different solutions 
to bring the light from the telescope to the CCL favored a full optical solution that 
includes prisms, mirrors and lenses.
The selected design uses 11 optical elements (see Fig.~\ref{figctrain}). The Coud\'e train 
takes the light beam with a prism at the Nasmyth-B platform and conduct it through 
the UT mechanical structure down to the Coud\'e room below each UT using a set 
of six prims and mirrors. 
The light is then routed from the UT Coud\'e room to the CCL, using two large lenses 
along the existing incoherent light ducts. 
The four Coud\'e trains relay a field of 17 arcsec around the acquired astronomical 
target to the CCL. The four beams from four UTs are combined in the CCL, where mode
selection and beam conditioning is performed by the 
fore-optics of the Front-End subsystem. 
All four Coud\'e Trains provide seeing limited images to the front end that 
is installed at the CCL.

\subsection{The Front-End}

\begin{figure}
\includegraphics[width=11.5cm,angle=0]{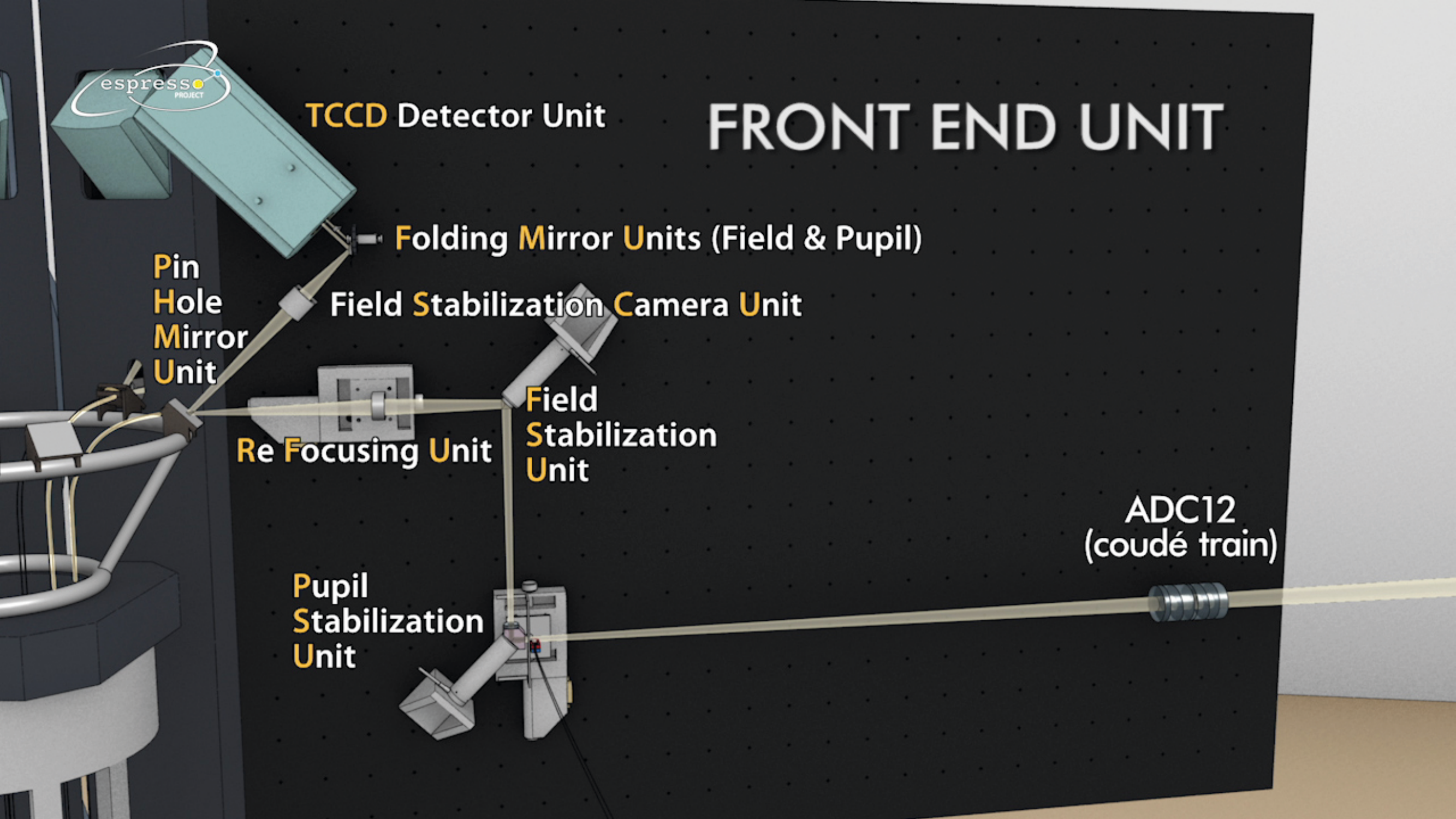}
\caption{Front-End unit and the arrival of one UT beam at the CCL. 
The same is replicated for the other UTs. 
}
\label{figfendu}
\end{figure}

The Front-End conducts the light beam received at the CCL after correcting it for
atmospheric dispersion with the ADC to the common focal plane where the 
spectrograph fiber feeds are located. 
A toggling mechanism handles the selection between the possible observational
modes in a fully passive way.
The beam conditioning is performed applying pupil and field stabilization 
(see Fig.~\ref{figfendu}). 
These are achieved via two independent control loops
each consisting of a technical camera and a tip-tilt stage. Another dedicated stage 
delivers a focusing function. The Front-End also handles the injection of the calibration
light from the calibration unit into the fibers and then into the spectrograph. 
A laser frequency comb (LFC) system is foreseen as main calibration source.
It produces a regular spectrum of lines equally spaced in frequency with an 
accuracy and stability linked to an atomic clock. 
The short-term Doppler shift repeatability of the LFC system has been tested 
in HARPS spectrograph and demonstrated to achieve the \cms 
level~\citep{wil12,pro16,pro20}.
The required repeatability of the order of ${\Delta \lambda}/{\lambda} \approx 10^{-10}$
cannot be guaranteed with currently used spectral sources such as thorium argon
spectral lamps, iodine cells, etc, but can be obtained with a LFC that would provide 
a spectrum sufficiently wide, rich, stable and uniform for this 
purpose~\citep{loc12,mol13lfc,gon20}. 
However, the long-term stability and reliability of a LFC system has not been proved yet. 
We refer to~\citet{sch21} for a discussion about the status of the current 
ESPRESSO calibration system.
However, two ThAr lamps for both simultaneous reference and calibration are also 
available as backup calibration sources, together with one simultaneous stabilized 
Fabry-P\'erot unit, also as a backup solution.

\subsection{The Fiber-Link}

The Fiber-Link sybsystem transfers the light from the Front-End to the spectrograph and
creates the a pseudo-slit in the output end inside the vacuum vessel. The 1-UT mode
uses a bundle of two octogonal fibers each, one for the astronomical object and one for
the sky or simultaneous reference. In the high-resolution mode (singleHR) mode, the fiber
has a core of 140 $\mu$m, equivalent to 1\arcsec ~on the sky; in the ultra-high 
resolution (singleUHR) mode the fiber core is 70~$\mu$m, covering a field of view 
of 0.5\arcsec.
The fiber entrances are organized in pickup heads that are moved to the focal plane 
of the Front End when the specific bundle of the specific mode is selected. 
In the 4-UT mode (multiMR) four object fibers and four sky/reference fibers are fed 
simultaneously by the four telescopes. 
The four object fibers and the four sky/reference fibers finally feed two 
separate single square fibers of 280 $\mu$m, for the object and for the sky/reference, 
respectively. 
In the 4-UT mode the spectrograph also {\it sees} a pseudo slit of four fiber 
square images twice as wide as the 1-UT fibers. 
One essential task performed by the Fiber-link subsystem is the light scrambling. 
The use of a double-scrambling optical system ensures both scrambling of the 
near field and far field of the light beam. 
A high scrambling gain, which is crucial to obtain the required RV precision in 
the 1-UT mode is achieved by the use of octagonal fibers \citep{cha12}.
The fiber link was upgraded in June 2019, thus increasing the photon-detection 
efficiency reaching more than 10\% at seeing better than 
0.75\arcsec ~\citep[see Fig.~12 in][]{pep21}.

\begin{figure}
\includegraphics[width=11.5cm,angle=0]{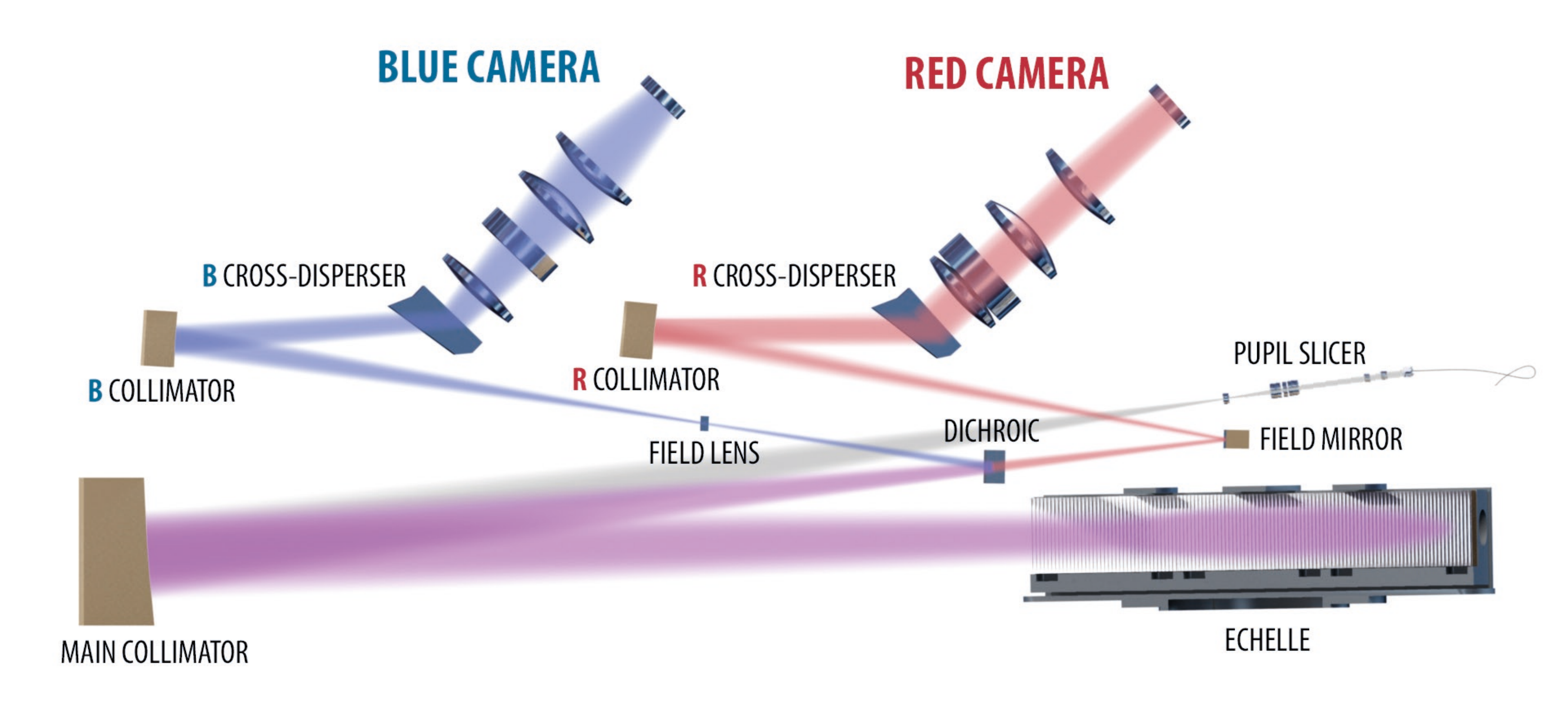}
\caption{Layout of the ESPRESSO spectrograph and its optical elements.
}
\label{figoptics}
\end{figure}

\subsection{Optical Design}

Designing a high efficiency and high resolution spectrograph is not an easy task due to
the large mirrors of the VLT telescopes and the 1 arcsec aperture of the instrument.
In order to minimize the size of the optics, and in particular, that of the main 
collimator and \'echelle grating, ESPRESSO implements an anamorphic optics, the APSU, 
which compresses the size of the pupil in the direction of the cross-dispersion. 
The pupil is then sliced in two by a pupil slicer and the slices are overlapped on
the \'echelle grating, leading to a doubled spectrum on the detector. 
This design reduces significantly the sizes of the optics and the \'echelle grating.
Without this {\it trick}, the collimator beam size would have been 40 cm in diameter 
and the size of the \'echelle grating would have reached a size of $240 {\times} 40$~cm. 
The size of current \'echelle grating of ESPRESSO is {\it only} $120 {\times} 20$~cm
and this also allows the use of much smaller optics (collimators, cross dispersers, 
etc.). The \'echelle grating is an R4 Echelle of 31.6 l\,mm$^{-1}$ and a 
blaze angle of 76$^{\circ}$. 
This solution significantly reduces the overall costs. 
The drawback is that each spectral element is covered by more detector pixels 
given the doubled image of the object fiber and its elongated shape on the CCD. 
In order to avoid to increase the detector noise, heavy binning is done in the 
case of faint-object observations, especially in the 4-UT mode.

The main components of the optical design are (see Fig.~\ref{figoptics}):

\begin{itemize}

\item{{\it The Anamorphic Pupil Slicing Unit (APSU).}  At the spectrograph entrance 
the APSU shapes the beam in order to compress it in cross-dispersion and splits 
in two smaller beams, while superimposing them on the \'echelle grating to minimize 
its size. The rectangular white pupil is then re-imaged and compressed.}

\item{{\it Dichroic.} Given the wide spectral range, a dichroic beam splitter separates 
the beam in a blue and a red arm, which in turn allows to optimize each arm for 
image quality and optical efficiency.}

\item{{\it Volume Phase Holographic Gratings (VPHGs).} The cross-disperser enables 
to  separate the dispersed spectrum in all its spectral orders. In addition, an
anamorphism is re-introduced to make the pupil square and to compress the order 
height such that the inter-order space and the SNR per pixel are both maximized. 
Both functions are accomplished using Volume Phase Holographic Gratings (VPHGs) 
mounted on prisms.} 

\item{{\it Fast Cameras.} Two optimised camera lens systems image the full
spectrum from 380 nm to 780 nm on two large $92{\times}92$ mm CCDs with 
10-$\mu$m pixels.}

\end{itemize}

\begin{figure}
\includegraphics[width=11.5cm,angle=0]{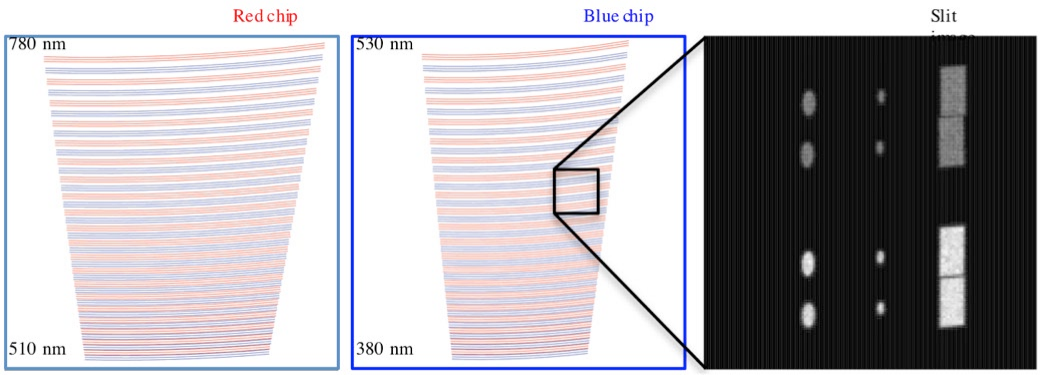}
\caption{Spectral format of the red (left panel) and blue (middle panel) spectra,
and a zoom of the pseudo-slit (right panel), showing the image of the target (bottom)
and sky (top) fibers. Each fiber is re-imaged into two slices. The three sets of fibers,
corresponding (from left to right) to the standard resolution 1-UT mode, ultra-high 
resolution 1-UT mode and the mid-resolution 4-UT mode (shown simultaneously for 
comparison).
This figure has been taken from \citet{pep14AN}.
}
\label{figslit}
\end{figure}

A sketch of the optical layout is depicted in Fig.~\ref{figoptics}. The spectral
format covered by the blue and the red chips as well as the shape of the pseudo
slit are displayed in Fig.~\ref{figslit}. 
In order to precisely compute the relative Earth motion to be able to properly correct
the RV measurement, it is necessary to calculate the weighted mean time of exposure.
Thus, the spectrograph is also equipped with an advanced exposure meter that 
measures the flux entering the spectrograph as a function of time. 
Its innovative design (based on a simple diffraction grating) allows a flux measurement 
and an RV correction at different spectral channels, in order to cope with possible 
chromatic effects that could occur during the scientific exposures. 
The use of various channels also provides a redundant and thus more reliable 
evaluation of the mean time of exposure.

\subsection{The opto-mechanics}

\begin{figure}
\includegraphics[width=11.5cm,angle=0]{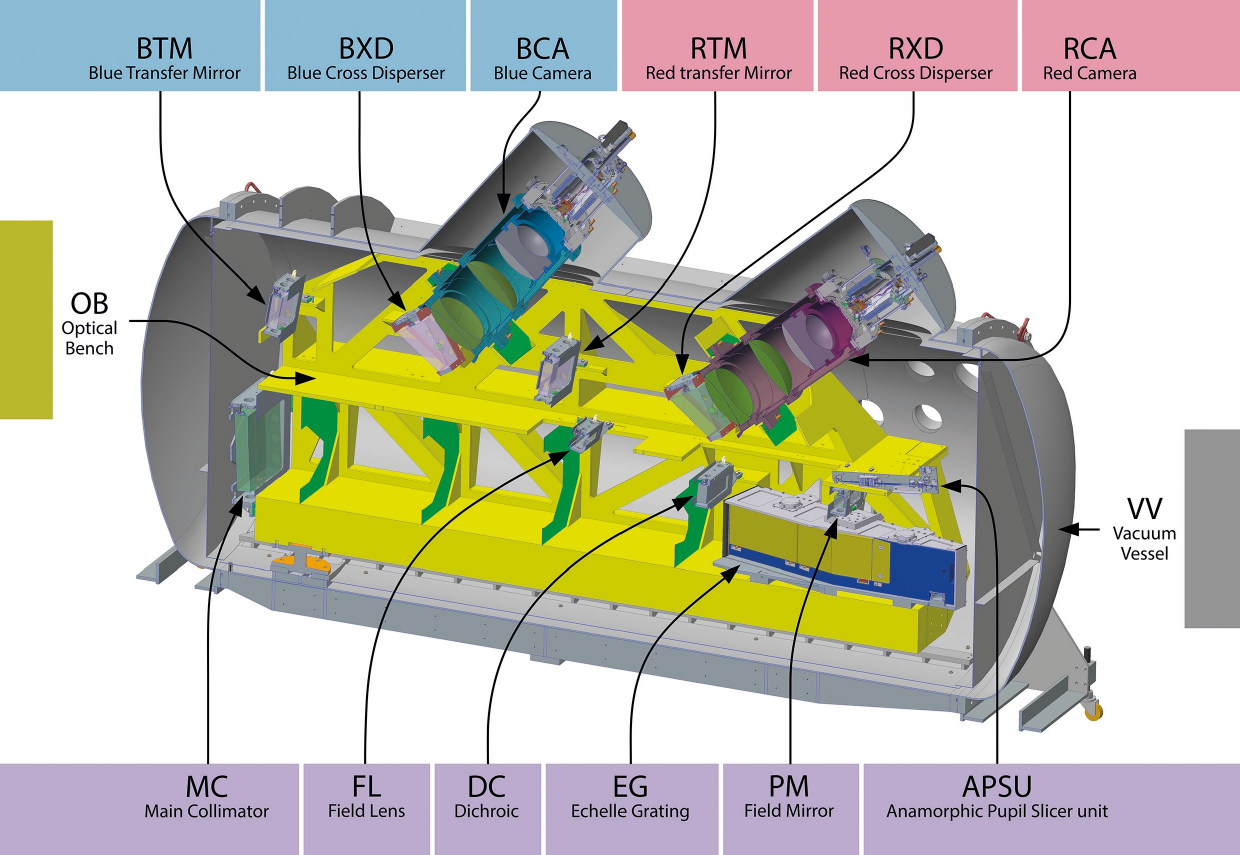}
\caption{Opto-mechanics of the ESPRESSO spectrograph.
}
\label{figomech}
\end{figure}

\begin{figure}
\includegraphics[width=11.5cm,angle=0]{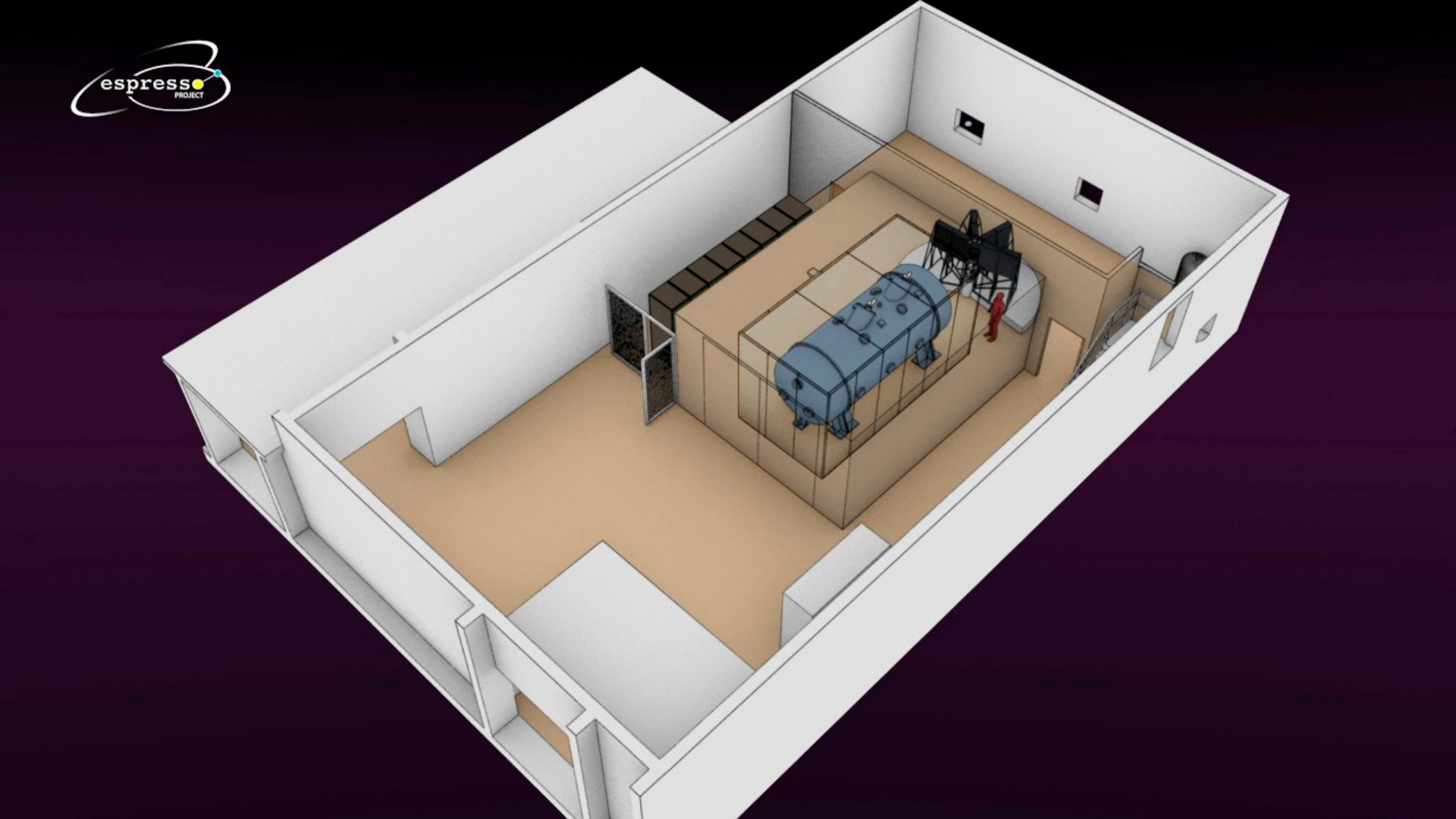}
\caption{Schematic view of the Combined Coud\'e Laboratory where ESPRESSO 
spectrogragh is located inside the vacuum vessel, and several thermal enclosures 
in a multi-shell control system.
}
\label{figvacves}
\end{figure}

ESPRESSO has been designed to be an ultra-stable spectrograph enabling RV
precisions of the order of 10 \cms, i.e. one order of magnitude better than its
predecessor HARPS. ESPRESSO is therefore built with a totally fixed
configuration and with the highest thermo-mechanical stability. 
The spectrograph optics are mounted in an optical bench specifically 
designed to keep the optical system within the thermo-mechanical tolerances 
required for high-precision RV measurements. 
The bench is mounted in a vacuum vessel in which 10$^{-5}$ mbar class vacuum 
is maintained during the entire duty cycle of the instrument. 
An overview of the opto-mechanics is shown in Fig.~\ref{figomech}. 
The temperature at the level of the optical system is required to be stable at the
mK level in order to avoid both short-term drift and long-term mechanical
instabilities. 
Such an ambitious requirement is obtained by locating the spectrograph in a 
multi-shell active thermal enclosure system as shown in Fig.~\ref{figvacves}. 
Each shell improves the temperature stability by a factor 10, thus getting from 
typically Kelvin-level variations in the CCL down to 0.001 K stability inside the 
vacuum vessel and on the optical bench.

\subsection{Large-area CCDs}

The CCDs are another innovative solution in the ESPRESSO project.
Large monolitic state-of-the-art CCDs have been chosen to use the optical 
field of ESPRESSO and to further improve the stability compared to the mosaic solution
employed in HARPS.
The sensitive area of the e2v chip is $92 {\times} 92$ mm covering 8.46$\times 10^{7}$ 
pixels of 10 $\mu$m size. Fast read out of such a large chip is achieved by using its 
16 output ports  at high speed. Other requirements on CCDs are very demanding, 
e.g. in terms of Charge Transfer Efficiency (CTE) and all the other parameters 
affecting the definition of the pixel position, immediately reflected into the 
radial-velocity precision and accuracy. 
The RV precision of 10 cm s$^{-1}$  rms requires measuring spectral line position 
changes of 2 nm (physical) in the CCD plane, equivalent to only 4 times the silicon 
lattice constant.
For better stability and thermal-expansion matching the CCD package is made of silicon 
carbide. 
The package of the CCDs, the surrounding mechanics and precision temperature control 
inside the cryostat head and its cooling system, as well as the thermal stability and 
the homogeneous dissipation of the heat locally produced in the CCDs during operation 
are of critical importance. ESO has thus built a new {\it superstable} cryostat that 
has already demonstrated excellent short-term stability. 

\section{Data flow system}

The ESPRESSO project has the final goal to provide the user with scientific data as 
complete and precise as possible in a short time (within minutes) after the end of an 
observation, to increase the overall efficiency and the ESPRESSO scientific output. 
For this purpose a software-cycle integrated view, from the observation preparation 
through instrument operations and control to the data reduction and analysis has been 
adopted.
Coupled with a careful design this ensures optimal compatibility and  
facilitates the operations and maintenance within the existing ESO Paranal Data Flow 
environment both in service and visitor mode. ESPRESSO Data Flow System
presents the following main subsystems:

\begin{itemize}
 
\item{\it  The ESPRESSO Observation Preparation Software (EOPS)}: a dedicated 
visitor tool (able to communicate directly with the VOT - Visitor Observing Tool) to 
help the observer to prepare and schedule ESPRESSO observations at the telescope 
according to the needs of planet-search surveys or other scientific programs.
The tool allows users to choose the targets best suited for a given night and to 
adjust the observation parameters in order to obtain the best possible quality of data.

\item{\it The Data Reduction Software (DRS)}: ESPRESSO has a fully automatic 
data reduction pipeline with the specific aim of delivering to the user high-quality 
reduced data, science ready, already in a short time after an observation has been 
performed. 
The computation of the RV at a precision reaching the 10~\cms\ level is an 
integral part of the DRS. 
During the  GTO, the ESPRESSO consortium has developed two codes to
improve the RV computation: (i) an automatic model-based telluric 
correction code~\footnote{https://github.com/RomainAllart/Telluric\_correction} 
for the ESPRESSO data reduction software~\citep{all22}; (ii) semi-Bayesian 
precise radial velocity computation code~\footnote{https://github.com/iastro-pt/sBART} 
through template matching~\citep{sil22},
both enabling for instance to achieve the 14, 8 and 10~\cms precision in 
M-, K-, and G-type stars, respectively~\citep{sil22}.
Coupled with the need to optimally remove the instrument signature, to take account 
the complex spectral and multi-HDU FITS format, the handling of the simultaneous 
reference technique and the multi-UT mode makes the DRS a truly challenging 
component of the DFS chain.

\item{\it The Data Analysis Software (DAS)}: dedicated data analysis software  
allows to obtain the best scientific results from the observations directly at the 
telescope (see e.g. Fig.~\ref{figspectcet}).
A robust package of recipes tailored to ESPRESSO, taking full advantage of the 
existing ESO tools (based on CPL and fully compatible with Reflex), addresses 
the most important science cases for ESPRESSO by analyzing (as automatically as 
possible) spectra of stars and quasars (among others, tasks such as 
line Voigt-profile fitting, estimation of stellar atmospheric parameters, 
normalization of stellar spectra and comparison with synthetic spectra, 
quasar continuum fitting, identification of absorption systems).

\item{\it Templates and control}: compared to other standalone instruments, the
main reason for the complexity of the ESPRESSO acquisition and observation 
templates are the possible usage of any combination of UTs, besides the proper 
handling of the simultaneous reference technique. 
ESPRESSO will contribute to open the new path for the control systems of future 
ESO instrumentation. 

\end{itemize}

\subsection{End-to-end operation}
ESPRESSO combine an unprecedent RV and spectroscopic precision with
the largest photon collecting area available today at the European Southern 
Observatory, and with an unique resolving power up to $R\sim200,000$.
In the singleHR mode, ESPRESSO can be fed with the light of an astromomical
object coming from any of the four 8.2m VLT telescopes, 
which significantly improves the scheduling flexibility for ESPRESSO 
programmes and surely will optimize the use of VLT time. 
The singleHR mode operates at a resolution of $R \sim 140,000$ with a RV precision 
of 10 \cms, opening the possibility to explore a new population of rocky planets orbiting
the habitable zones of solar-type stars.
The scheduling flexibility is a fundamental advantage for survey programmes like RV 
searches for exoplanets or time-critical programmes like studies of transiting 
planets. 
The 1-UT mode also offers the possibility to enhance the resolving power up to 
200,000 with a extremely stable wavelength accuracy which certainly will motivate 
new scientific projects e.g. high-accuracy stellar astrophysics projects.
In addition, in the multiMR mode, ESPRESSO is able to collect the light of up 
to four UTs to generate a high-resolution spectrum. 
The effectively larger telescope aperture of about 16m provides access to faint 
astronomical targets at a resolution of $R \sim 70,000$.
ESPRESSO shall not be considered as a stand-alone instrument but as a 
science-generating machine, certainly delivering full-quality scientific data in 
less than a minute after the end of an observation. 


\begin{acknowledgement}
The ESPRESSO project is supported by the Swiss National Science Foundation 
program FLARE, Italian Institute of Astrophysics (INAF), Instituto de Astrof{\'\i}sica
de Canarias (IAC, Spain), Instituto de Astrof{\'\i}sica e Ci\^encias do 
Espa\c{c}o/Universidade de Porto and Universidade de Lisboa (Portugal), 
and the European Southern Observatory (ESO).
J.I.G.H. acknowledges financial support from the Spanish Ministry of Science and 
Innovation (MICINN) project PID2020-117493GB-I00 and from the Government of the 
Canary Islands project ProID2020010129.
F.P. would like to acknowledge the Swiss National Science Foundation (SNSF) for 
supporting research with ESPRESSO through the SNSF grants nr. 140649, 152721, 
166227, 184618 and 215190. The ESPRESSO Instrument Project was partially funded 
through SNSF’s FLARE Programme for large infrastructures.
N.C.S. acknowledges the support by FCT - Fundação para a Ciência e a Tecnologia 
through national funds and by FEDER through COMPETE2020 - Programa Operacional 
Competitividade e Internacionalização by these grants: UIDB/04434/2020; UIDP/04434/2020. 
Co-funded by the European Union (ERC, FIERCE, 101052347). 
Views and opinions expressed are however those of the author(s) only and do not 
necessarily reflect those of the European Union or the European Research Council. 
Neither the European Union nor the granting authority can be held responsible for them.
The authors wish to acknowledge the exceptional work and enthusiasm delivered by 
all the members of the ESPRESSO team and warmly thank them for significantly 
contributing to the successful completion of the project.
\end{acknowledgement}

\bibliographystyle{spbasicHBexo}  
\bibliography{gonzalezhernandez_update} 

\end{document}